# Nanotwinning in boron subphosphide $B_{12}P_2$

B.A. Kulnitskiy,[a,b] I.A. Perezhogin,[a,b,c] V.D. Blank,[a,b] V.A. Mukhanov,[d] and V.L. Solozhenko[d,*]

[a] *Technological Institute for Superhard and Novel Carbon Materials, 108840 Troitsk, Russia*
[b] *Moscow Institute of Physics and Technology, 141700 Dolgoprudny, Russia*
[c] *International Laser Center, Lomonosov Moscow State University, 119991 Moscow, Russia*
[d] *LSPM–CNRS, Université Paris Nord, 93430 Villetaneuse, France*

Microstructure of boron subphosphide $B_{12}P_2$ produced by self-propagated high-temperature synthesis has been studied by high-resolution transmission electron microscopy. Two systems of twins have been found i.e. conventional twins on the $(0003)_h$ plane and nanotwins resulting from duplication of the rhombohedral unit cell of $B_{12}P_2$ along one of the basic vectors.

Boron subphosphide $B_{12}P_2$ is a hard refractory wide bandgap semiconductor with outstanding high-temperature stability [1-4] and a promising material for a wide range of engineering applications [5]. Here we report the results of transmission electron microscopy studies of $B_{12}P_2$ produced by the recently developed method of self-propagated high-temperature synthesis that is characterized by the simplicity of implementation, high efficiency, low cost of the product, and good perspectives for large-scale production [6].

Microcrystalline powder of boron subphosphide has been synthesized by self-propagating high-temperature reaction of boron phosphate, magnesium diboride and metallic magnesium according to the method described in detail elsewhere [6]. According to X-ray diffraction study (TEXT 3000 Inel, CuK$\alpha$1 radiation) the sample is well-crystallized single-phase $B_{12}P_2$ with lattice parameters $a = 0.5992(4)$ nm, $c = 1.1861(8)$ nm that are close to the literature data [7].

Microstructure of boron subphosphide has been studied by high-resolution transmission electron microscopy (HRTEM) using JEM-2010 microscope. According to TEM data, the $B_{12}P_2$ powder consists of thin flat hexagonal particles having dimensions from tens to several hundred nanometers; and a significant part of the particles contains twins.

A characteristic HRTEM image of boron subphosphide particle is presented in Fig. 1a. The right part of the image is the perfectly ordered $B_{12}P_2$ structure without twins and stacking faults, while the left part clearly demonstrates the nanotwinned structure similar to that of $B_6O$ [8]. The corresponding fast Fourier transform (FFT) images are shown in Fig. 1b and Fig. 1c. One can see that Fourier image of nanotwinned structure (Fig. 1b) contains extra spots caused by superstructure e.g. an additional spot corresponding to the 0.948 nm interplanar distance appears in the middle of the radius vector to the $(0\text{-}111)_h$ spot, which corresponds to the 0.474 nm interplanar distance of the $B_{12}P_2$ crystal lattice.

---

[*] vladimir.solozhenko@univ-paris13.fr

The observed superstructure can be formed by the duplication of rhombohedral unit cell of pristine $B_{12}P_2$ along one of the basic vectors. Fragment of the $B_{12}P_2$ crystal lattice which corresponds to a nanotwin is presented in Fig. 2. The new lattice can be formally described as a triclinic one with the following parameters: $a = b = 5.24$ Å, $c = 10.48$ Å; $\alpha = \beta = \gamma = 69.6°$, and allow us to explain the appearance of all additional spots in the Fourier image of nanotwinned structure (Fig. 1b). The (001) plane of the new triclinic lattice is parallel to the $(01\text{-}11)_h$ (or $(100)_r$) plane of the crystal lattice of pristine $B_{12}P_2$. Such nanotwins (Fig. 1a) can also be considered as a result of the structural (rhombohedral to triclinic) transformation in $B_{12}P_2$ crystals. The presence of long streaks passing through spots in the Fourier images (Fig. 1b,c) is most likely due to the fine lamellar structure of the nanotwin fragments. It should be noted that experimentally observed structure of nanotwins in boron subphosphide completely differs from the nanotwinned structure theoretically predicted for $B_{12}P_2$ [9].

In addition to the nanotwins described above, the conventional twinning was also observed. Fig. 3 shows the twin structure and corresponding Fourier image (the twin reflections are circled). The twin boundary is $(0003)_h$ plane (or (111) in rhombohedral coordinates). The same twinning plane was found for $B_{12}As_2$ [10], however, in general $(0003)_h$ twinning boundary is not typical for the boron-rich compounds with structure related to α-rhombohedral boron.

Both nanotwins and conventional twins in $B_{12}P_2$ formed in the course of ultrafast (a few milliseconds) self-propagated high-temperature (> 3000 K) reaction may significantly improve the material strength by blocking dislocation movements which open new possibilities for creation and design of advanced materials based on boron subphosphide.

V.A.M. and V.L.S. acknowledge financial support from the European Union's Horizon 2020 Research and Innovation Programme under the Flintstone2020 project (grant agreement No 689279).

**References**

1. Solozhenko, V.L., Bushlya, V. Mechanical properties of boron phosphides. *J. Superhard Mater.*, 2019, vol. 41, No. 2, pp. xx-xx.

2. Solozhenko, V.L., Mukhanov, V.A., Sokolov, P.S., Le Godec, Y., Cherednichenko, K.A., Konôpková, Z. Melting of $B_{12}P_2$ boron subphosphide under pressure. *High Press. Res.*, 2016, vol. 36, No. 2, pp. 91-96.

3. Solozhenko, V.L., Cherednichenko, K.A., Kurakevych, O.O. Thermoelastic equation of state of boron subphosphide $B_{12}P_2$. *J. Superhard Mater.*, 2017, vol. 39, No. 1, pp. 71-74.

4. Reshetniak, V.V., Mavrin, B.N., Edgar, J.H., Whiteley, C.E., Medvedev, V.V. Phonon states of $B_{12}P_2$ crystals: Ab initio calculation and experiment. *J. Phys. Chem. Solids*, 2017, vol. 110, pp. 248-253.

5. Emin, D. Unusual properties of icosahedral boron-rich solids. *J. Solid State Chem.*, 2006, vol. 179, No. 9, pp. 2791-2798.

6. Mukhanov, V.A., Sokolov, P.S., Brinza, O., Vrel, D., Solozhenko, V.L. Self-propagating high-temperature synthesis of boron subphosphide $B_{12}P_2$. *J. Superhard Mater.*, 2014, vol. 36, No. 1, pp. 18-22.

7. Yang, P., Aselage, T.L. Synthesis and cell refinement for icosahedral boron phosphide $B_{12}P_2$. *Powder Diffr.*, 1995, vol. 10, No. 4, pp. 263-265.

8. An, Q., Reddy, K.M., Dong, H., Chen, M.-W. Oganov, A.R., Goddard, W.A. Nanotwinned boron suboxide ($B_6O$): New ground state of $B_6O$. *Nano Lett.*, 2016, vol. 16, No. 7, pp. 4236-4242.

9. An, Q., Goddard, W.A. Ductility in crystalline boron subphosphide ($B_{12}P_2$) for large strain indentation. *J. Phys. Chem. C*, 2017, vol. 121, No. 30, pp. 16644−16649.

10. Chen, H., Wang, G., Dudley, M., Zhang, L., Wu, L., Zhu, Y., Xu, Z., Edgar, J.H., Kuball, M. Defect structures in $B_{12}As_2$ epitaxial layers grown on (0001) 6*H*-SiC, *J. Appl. Phys.*, 2008, vol. 103, No. 12, pp. 123508-1-9

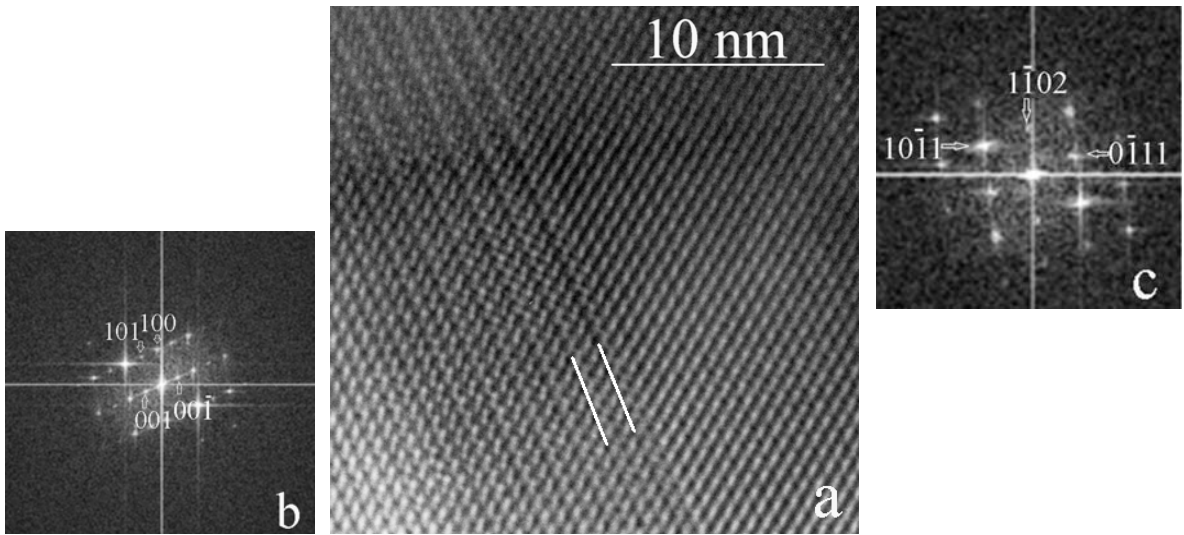

Fig. 1. (*a*) A representative HRTEM image displaying two $B_{12}P_2$ structures: the perfectly ordered one (right part) and the nanotwinned structure (left part). Fragment of a nanotwin is marked by white dashes. (*b*) FFT image of the left part of (*a*); (*c*) FFT image of the right part of (*a*) (the indices are given for the hexagonal system).

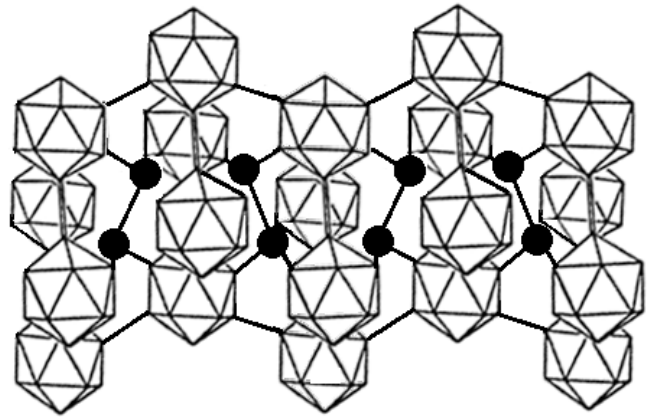

Fig. 2. Fragment of the $B_{12}P_2$ triclinic lattice which corresponds to a nanotwin. Icosahedra are the $B_{12}$ units, black circles show phosphorus atoms.

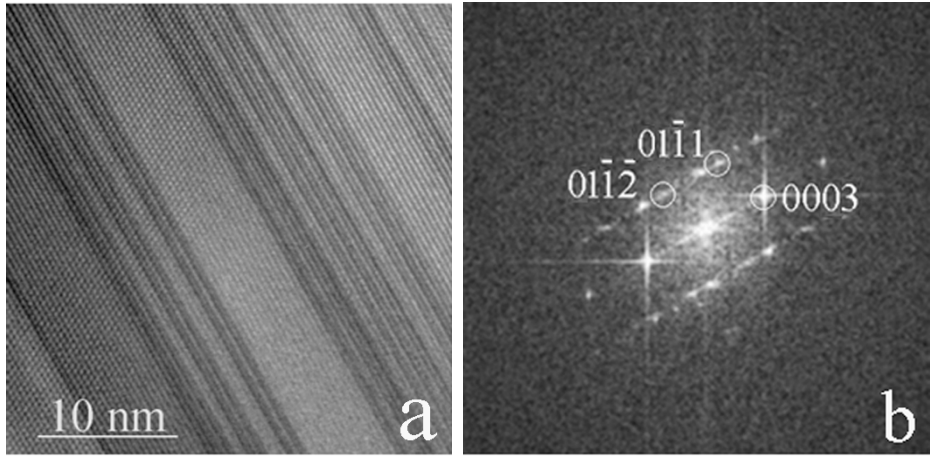

Fig. 3. HRTEM image displaying twins in $B_{12}P_2$ (*a*) and the corresponding FFT image (*b*) (the indices are given for the hexagonal system).